\documentclass[notitlepage, twocolumn, letterpaper, prx, longbibliography, 10pt]{revtex4-1}
\usepackage[utf8]{inputenc}
\usepackage{amsmath}
\usepackage{amssymb}
\usepackage{graphicx}
\usepackage{hyperref}
\usepackage{subcaption}
\usepackage[percent]{overpic}
\usepackage{pdfpages}
\usepackage[normalem]{ulem}
\usepackage[switch,columnwise]{lineno}
\usepackage[margin=1 in]{geometry}
\usepackage[object=vectorian]{pgfornament} 

\newcommand{\missingcite}[1]{\textcolor{green}{\textbf{[cite]}}}

\makeatletter
\AtBeginDocument{\let\LS@rot\@undefined}
\makeatother

\graphicspath{{images/}}

\begin{document}

\title{Data-Driven Modeling of U.S.~Ideological Dynamics}

\author{David Sabin-Miller}
\email{dasami@umich.edu}
\affiliation{Center for the Study of Complex Systems, University of Michigan, Ann Arbor, MI, USA}
\author{Christopher Harding}
\affiliation{Department of Physics, University of Michigan, Ann Arbor, MI, USA}




\begin{abstract}

    The dynamics of political opinion are a critical component of modern society with large-scale implications for the evolution of intra- and international political discourse and policy.
    Here we utilize recent high-resolution survey data to quantitatively capture leading-order psychological and information-environmental patterns. We then inform simulations of a theoretical dynamical framework with several different models for how populations' ideology evolves over time, including a plausible model which reproduces current macro-scale ideological distributions given the empirical micro-scale data gathered. 
    This effort represents an attempt to discover true underlying trends of political reasoning in general audiences, and to extrapolate the long-term 
    implications of those trends as they interact with the political exposure landscape. 
    Accurate modeling of this ecosystem has the potential to predict catastrophic outcomes such as hyperpolarization, and to inform effective intervention strategies aimed at preserving and rebuilding constructive political communication. 
\end{abstract}

\maketitle


\section{Introduction} \label{sec:intro}
The dynamics of popular ideology are immensely consequential for the political evolution of societies, particularly democracies. It is a matter of some concern, then, that modern information ecosystems allow individuals to select their favored sources and narratives according to their current beliefs and proclivities, which in turn impacts their ideological evolution, creating a feedback loop between ideology and political information received. Here we follow up on our previous work \cite{DSM2020} which proposed a dynamical framework linking these systems, by informing every aspect of that framework with real-world data, examining the consequences, and proposing future improvements to both theory and data.

This work contributes to a rich history of scholarship endeavoring to understand social and political systems through modeling. This includes a myriad of ``toy models" examining the implications of various individual-level mechanisms on macroscopic outcomes (e.g., \cite{galam1997rational, deffuant2000mixing, hegselmann2002opinion,  galam2004contrarian,Sood2005voter, galam2008sociophysics,hegselmann2005opinion, galam2007role, martins2008continuous, verma2014impact, Knopoff_2014, liu2014control, porter2016dynamical, wang2016bistability, DelVicario2017, pinasco2017modeling, sirbu2017opinion, kan2023adaptive} and others) and statistical-physics-based analogies (e.g., \cite{Castellano2009statmech, toscani2006kinetic,bertotti2008discrete,during2009boltzmann,during2015opinion}). We hope this work serves as a bridge to empirical and theoretical analyses of political media and its impact (e.g., \cite{groseclose2005measure,entman2007framing, Slater2007spiral, DellaVigna/Kaplan2006, Martin2017, acemoglu2011opinion, fan2017evolution}), and that this dynamical modeling mindset contributes to the social-science conversation around political polarization (e.g., \cite{dimock2014political, hare2014polarization, iyengar2015fear, pew2017divide, mason2018uncivil,gimpel2020urban}). We hope that this work serves as a useful marriage between the theoretical modeling community and the empirical analysis community, demonstrating the possibility of an iterative reality-seeking theory-experiment loop.



In a series of recent surveys (described in more detail in \cite{DSM2025spectrograph}) we collected empirical data with the goal of informing the theoretical framework for popular ideological dynamics in a one-dimensional ideology domain which was introduced in \cite{DSM2020}. The core hypothesis of that framework is that the largest-order determinant of individuals' reaction to political content is an internal sense of ``otherness"---a subjectively-felt distance in largely one-dimensional ideological space---likely manifesting as short-range attraction and long-range repulsion (with stochasticity standing in for additional situational nuance such as reasonableness and issue-based differences). The incoming stream of political stimuli (which we will call \textit{percepts}) is then represented as a biased random distribution according to the observer's current ideological position. We consider the possible psychological modifier of tribalism (i.e., favoring a concept more if it comes from one's ``in-group") by allowing observers and percepts to have a political party affiliation---though ideally future models could allow additional qualitative in-groups such as ethnicity, geographic region, or gender. 

In this paper we inform that framework with recently-gathered empirical results, generating systematically varying nonparametric distributions which replace formerly conjectured functional forms. However, due to the practical impossibility of measuring actual ideological \textit{movement} per statement, we must still conjecture plausible dynamical rules given the observable data of dissonance, agreement, and observer/statement party affiliations. It is in reconciling the observable low-level inputs with the observable high-level outcomes that our work as modelers lies.
Our simplest such model predicts that the U.S.~population should hyper-polarize given the late-2023/early-2024 information ecosystem, which may be true on long time scales, though we present additional data which cast doubt on this. However, seeking a model which produces current ideological distributions as equilibria, we augment the basic model with several reasonable effects, and achieve reasonable results without too much fine-tuning. However, this modeling leeway suggests that further experiments are required to outline more accurate ideological dynamical rules---fortunately, this approach offers a guideline for how such studies might quantitatively frame their explorations.


In this way we hope to kick-start a theory-experiment loop to allow models to more accurately model this vital societal system over time, predict catastrophic outcomes, and recommend strategies for interventions aimed at cultivating compromise and reconciliation. 


\section{Results}
\subsection{Data: Reactions (Agreement vs Dissonance)}
We conducted a survey in the Fall of 2023 ($N = 296$ participants, non-representative volunteers and Mechanical Turk Masters) and Spring of 2024 ($N = 508$ participants, age/sex/ethnicity/political representative sample through the Prolific survey platform) which aimed at examining consistent patterns of ideology as a variable, both internally and externally recognized \cite{DSM2025spectrograph}. Individuals' own ideology was estimated in multiple ways, and respondents then rated political opinion statements' ideological position and their reaction (level of agreement, and positive/negative emotional sentiment) to each statement. 
For more methodological details we refer to \cite{DSM2025spectrograph}.

\begin{figure}[tp]
    \center
    \includegraphics[width = \columnwidth]{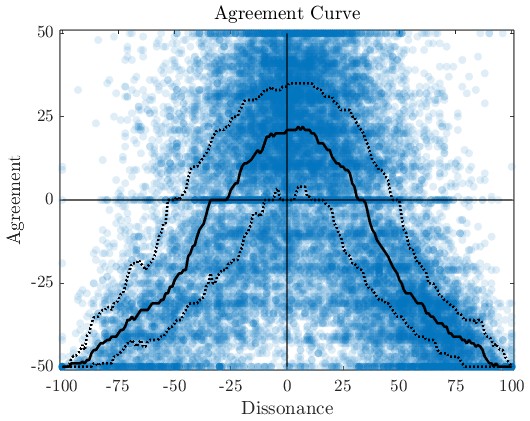}
    \caption{\textbf{Agreement Curve:} Agreement versus subjectively experienced dissonance for 23,736 statement-observer events. The black curve is a moving median (with window width of 7), with 25$^{th}$ and 75$^{th}$ percentiles (dotted). Horizontal striation occurs near multiples of ten (visible from survey software), not labels on the slider; there were seven labels per slider, roughly corresponding to positions $0, \pm16, \pm32,$ and $\pm48$.}
    \label{fig:reaction_scatter}
\end{figure}

\begin{figure}[tp] 
    \centering
    \includegraphics[width = .8\columnwidth]{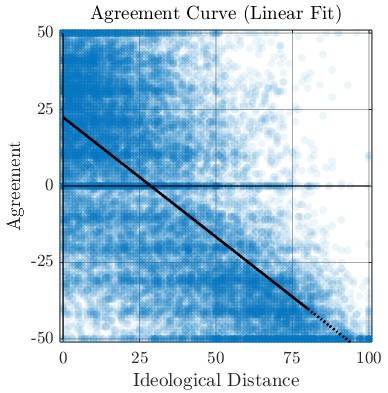}
    \caption{\textbf{Linear Agreement Model:} As Fig.~\ref{fig:reaction_scatter}, but with the data ``folded'' so that left and right halves are overlaid, and a least-squares linear fit: $a = 22.46 - 0.78d$. This fit appears to overestimate the ``spine" of the data for high distances, due to both edge effects and non-Gaussian-distributed errors. This fit was found by restricting the fitting data to $d<80$, which was the steepest result found for any such data-cutoff. 
    }
    \label{fig:reaction_scatter_folded}
\end{figure}

From these measurements we could calculate the signed distance in subjective ideological space (which we call the \textit{dissonance})---e.g., a person identifying as a -10 (somewhat liberal) person perceiving a statement they rate as +35 (strongly conservative) would experience a dissonance of +45. The primary objective of this survey was to explore the relationship between this dissonance and the observer's reaction to these political statements. The data from 23,436 observer-statement interactions is displayed in Figs.~\ref{fig:reaction_scatter} and \ref{fig:reaction_scatter_folded}. We of course find considerable vertical variance in agreement; some statements are more convincingly phrased than others, and different issues enjoy different patterns of acceptance, but in aggregate a clear macroscopic pattern emerges: the central $50\%$ of responses fall within a relatively narrow band (dotted 25/75$^{th}$ percentile curves in Fig.~\ref{fig:reaction_scatter}).

These curves are immediately of utility for those designing political messaging to build consensus. For instance, if one is attempting to reach an audience of a particular estimated ideology and wishes to have over 50\% of that audience agree with you (and all other things being equal), it's best to codify one's message so as to come across as falling within $\pm30$ of the audience on a $[-50,50]$ ideological scale (i.e., $\pm30\%$ of the political spectrum's width). For instance, to an audience of quite extreme ideologues at $-45$, a truly neutrally phrased message (i.e., coming across as 0 ideologically) is likely to engender primarily disagreement, while a more ideologically catered but moderate message of $-20$ would have a good chance of reaching them in a constructive way.  These quantities are of course subjective, and particularities of the issue and the audience may suggest additional strategies, but our analyses suggest a surprising amount of universality in the interpretation of this one-dimensional political spectrum across individuals regardless of ideology or party affiliation \cite{DSM2025spectrograph}.




As a simplest model of this trend, we may exploit the near-symmetry and seek a linear fit to the unsigned ideological distance, seen in Fig.~\ref{fig:reaction_scatter_folded}. To avoid the worst edge-effects in the vertical direction, we fit only to data with distance $<80$, but the fit still 
appears to have positive bias relative to the ``spine" of the data on the right-hand side---the significantly non-Gaussian spread of data makes this simple least-squares linear model unsatisfying.

\begin{figure}[tp] 
    \centering
    \includegraphics[width = \columnwidth]{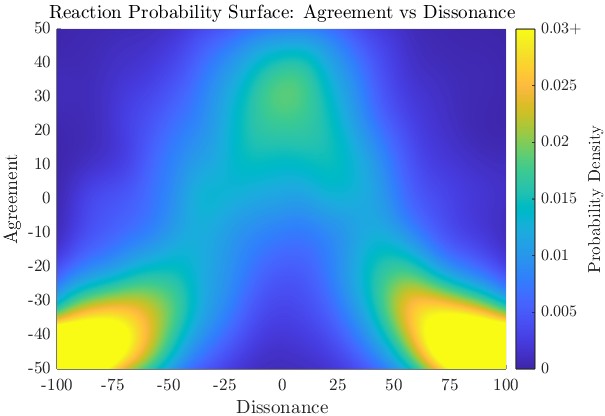}
    \caption{\textbf{Reaction Probability Surface:} 
    Heat-map view of a quasi-kernel-density surface generated by ``fuzzing" each data point from Fig.~\ref{fig:reaction_scatter} into a two-dimensional Gaussian distribution with standard deviations $\sigma_x = \sigma_y = 7$ in the $x$ and $y$ directions, to reflect the inherent uncertainty in each quantity, 
    then normalized so that each vertical slice yields a probability distribution of agreement for that level of dissonance. 
    }
    \label{fig:reaction_surface}
\end{figure}

However, for modeling purposes there is little reason to find such a trend and then add artificial ``noise" again (especially if that noise appears to be non-standard and asymmetric), when we can simply use the data itself and its inherent shape. For this purpose we adopt a quasi-kernel-density-estimate to construct a probability distribution which systematically varies with dissonance which is motivated by the inherent uncertainty of these measurements: we first ``fuzz'' each data point into a 2-dimensional Gaussian distribution, then normalize each vertical slice. The results are shown in Fig.~\ref{fig:reaction_surface}. 

The kernel width of $\sigma = 7$ in each dimension represents a reasonable estimate of the uncertainty of each particular response---in \cite{DSM2025spectrograph} we estimate an upper bound for uncertainty of an individual's own subjective ideological position to be $\sigma \leq 8$. However, further work is warranted to more precisely estimate the uncertainties of these particular quantities, agreement and dissonance
---such as by repeating some statements to see how much one individual's estimation of its ideological position, and their level of agreement with it, varies for any particular statement.

\begin{figure}[tp] 
    \centering
    \includegraphics[width = \columnwidth]{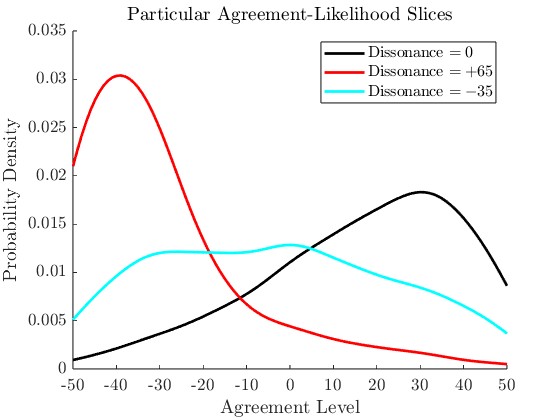}
    \caption{\textbf{Reaction Probability Distributions:}  Three particular vertical slices of Fig.~\ref{fig:reaction_surface}, showing the implied probability distributions for each level of dissonance on the $[-50,50]$ scale: 0 (in line with the observer's self-identified ideology), +65 (much more conservative than the observer) and -35 (significantly more liberal, just past the uncertainty boundary of $\approx\pm 30$).}
    \label{fig:reaction_particular_slices}
\end{figure}

This kernel method has the benefit of naturally providing the appropriate noise shape for each dissonance level instead of an artificially-Gaussian distribution. In Fig.~\ref{fig:reaction_particular_slices} we demonstrate the implied reaction probability distribution for a generic individual experiencing a dissonance of 0, +65, or -35, reading off corresponding vertical slices of the surface in Fig.~\ref{fig:reaction_surface}.

\begin{figure*}[tp] 
    \centering
    \begin{overpic}[width = \columnwidth]{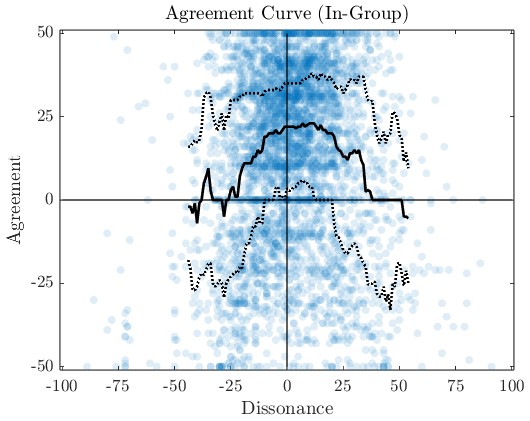} 
    \put(12, 76){\textbf{a}}
    \end{overpic}
    \begin{overpic}[width = \columnwidth]{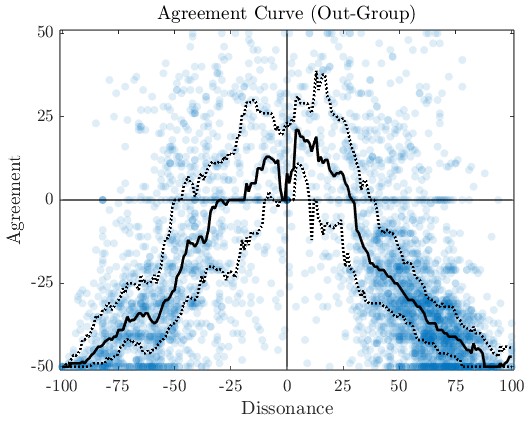} 
    \put(12, 76){\textbf{b}}
    \end{overpic}
    \caption{\textbf{Difficulty of quantifying tribalism:} Moving-median agreement curves for \textbf{a)} in-group percepts ($N = 4054$) and \textbf{b)} out-group percepts ($N= 4025$). Where the window contains fewer than 20 data points, the median and percentile curves are omitted. We see the challenge of isolating the effect of tribalism, due to the strong correlation between ideology and party: in-group statements tended to be low-dissonance and out-group statements the opposite. 
    In addition, sample sizes are significantly smaller due to subdividing: participants must be assigned the marked condition (50\%), identify as Strong or Lean towards one of the parties (77\%), and view a statement from one of the non-Centrist statement pools ($30/68 = 44\%$ each), resulting in the use of only about 17\% of our original dataset to inform each panel. }
    \label{fig:reaction_curve_ingroup_outgroup}
\end{figure*}

To investigate the effect of tribalism (in-group/out-group bias) on reactions, these data were collected under one of two randomly-assigned conditions, which differed only in whether statements were prefaced with their speaker's supposed  political party affiliation (E.g., ``A Democrat says \ldots" before each statement from the ``liberal" statement pool). However, as shown in Fig.~\ref{fig:reaction_curve_ingroup_outgroup} it is difficult to compare ``in-group" reactions and ``out-group" ones to draw any conclusions on the effect of explicitly advertised speaker identity, for several reasons. First, the resulting data for each type of statement/observer pair don't overlap much---out-group statements occupy primarily high-dissonance portions of the graph, while in-group ones fill out the low-dissonance portion. This is likely due to our statement pool's relative lack of ambiguous statements---as we found in \cite{DSM2025spectrograph}, the mean assessment of marked and unmarked statements were not significantly different. A dedicated experiment quantitatively exploring this effect (possibly including other in-group markers such as race, religion, socioeconomic status, culture, or fandom, which would be less confounded by heavy correlation with ideology) could elucidate this likely-important psychological bias.

\subsection{Data: Exposure (Percepts vs Ideology)}
We were able to construct political-content exposure surfaces which, given a respondent's ideology, provide us with an estimated probability distribution of their exposure to far-left, mid-left, centrist, mid-right, and far-right content. To aid incorporation of possible in-group/out-group interpretation bias, each of these ideological leanings was further broken down by which of three political party identities (Democrat-aligned,  Republican-aligned, or Independent/unaligned) the respondent considered their sources to hold, leading to breaking their total political information exposure into fifteen categories (e.g.~one might estimate that 20\% of their political diet comes from mid-left Democrat-aligned sources, etc.). 

These data are significantly more complex than individual statement ratings/reactions, and their accuracy from self-report is uncertain, since it is difficult to estimate one's entire sphere of political influences. Additional studies could estimate this more accurately in the future and be used in place of this relatively coarse self-reported data, but these do have inherent value as a subjective sense of what looms large in respondents' memories, with some clearly visible macroscopic patterns. 

\begin{figure}[htp] 
    \centering
    \includegraphics[width = \columnwidth]{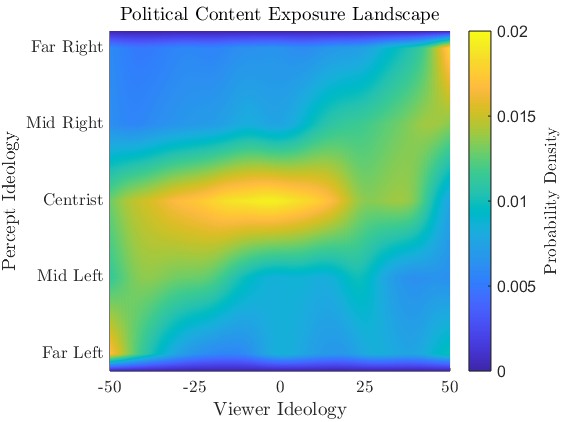}
    \caption{\textbf{Information ecosystem (single):} Heat-map view of average exposure frequency to each ideology of content ($y$ axis, with ``far right/left" identified with $\pm 45$ and ``mid right/left" identified with $\pm 22.5$) for observers of each ideology ($x$ axis). A vertical slice in this landscape represents the average distribution/``diet" of political content for individuals at that ideology value.}
    \label{fig:exposure_landscape_combined}
\end{figure}

\begin{figure*}[tp] 
    \centering
    \begin{overpic}[width = \textwidth]{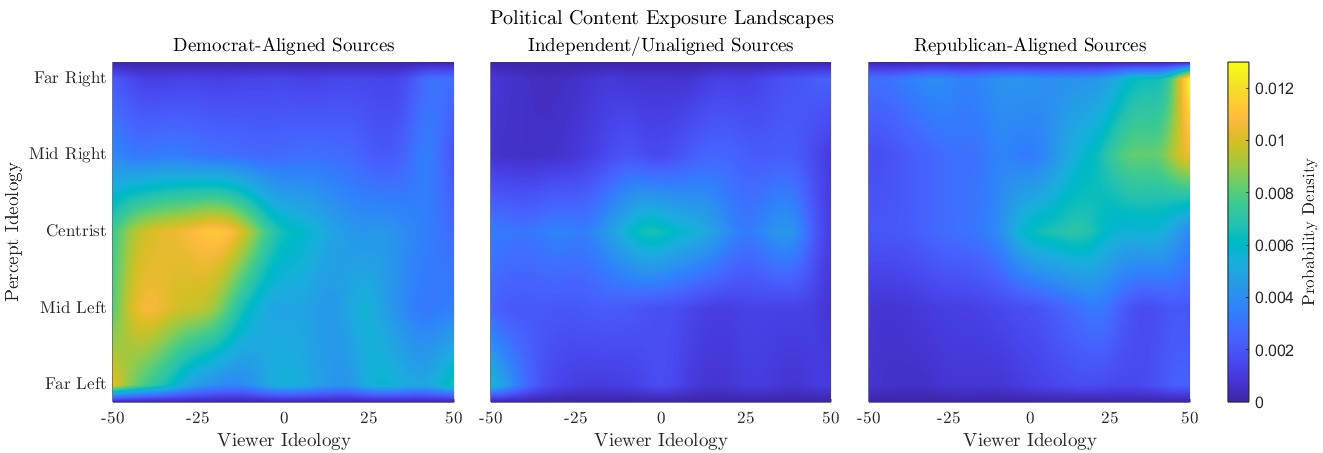}
    \put(8,30){\textbf{a}}
    \put(36,30){\textbf{b}}
    \put(65,30){\textbf{c}}
    \end{overpic}
    \caption{\textbf{Information ecosystem, by perceived source allegiance:} As Fig.~\ref{fig:exposure_landscape_combined}, but broken down by political affiliation of source (panels) for use in incorporating tribalism (i.e.~a differing impact depending on partisan allegiance). Given an observer at ideology $I$, this allows us to estimate their diet of each type and ideology of content by taking the corresponding slice of each surface (see Fig.~\ref{fig:exposure_distribs_particular}). 
    }
    \label{fig:exposure_landscape_triple}
\end{figure*}

We took a Gaussian-weighted moving average of kernel width $\sigma=5$ in the horizontal direction to estimate the average amounts of ``far left," ``mid-left," ``centrist," ``mid-right," and ``far right" content seen by observers at each ideology value. To translate those coarse qualitative terms to our quantitative ideology domain, a choice was made to equate them with $-45, -22.5, 0, 22.5,$ and $45$, and 
results were linearly interpolated between those five estimated ideological values vertically (i.e., between those ideology values). Each vertical slice was then normalized such that taking the slices corresponding to a particular observer-ideology results in three distributions which together add to an area of 1. The results are seen in Figs.~\ref{fig:exposure_landscape_combined} (combining all ``party allegiances" of sources) and \ref{fig:exposure_landscape_triple} (trusting respondents' assessment of sources' party allegiance).

In a model devoid of tribalism (i.e.~differing impact depending on a source's party allegiance, which our data indeed failed to precisely measure), we can use Fig.~\ref{fig:exposure_landscape_combined} to estimate the subjectively-experienced ``slice" of the political world that an average member of any particular ideology is exposed to. We see a broadly diagonal pattern, reflecting the unsurprising leading-order effect of self-serving/comfort-zone consumption bias. It is perhaps surprising, however, to see the spikes at the corners, where (a considerable number of) respondents were willing to not only identify as ``extremely liberal/conservative" ideologically but that these individuals also acknowledge that their political sources are also mostly ``far left/far right."

However, we did ask respondents to distinguish the political affiliation of their sources, and Fig.~\ref{fig:exposure_landscape_triple} shows the more granular results. We see a primary asymmetry emerge between the left and right panels: liberal respondents (whose diets are represented by the left halves of all three panels) report consuming primarily Democrat-aligned mid-left and centrist content \textit{above} the ideological diagonal i.e.~more centrist than themselves (bright triangle in Fig.~\ref{fig:exposure_landscape_triple}a), but very little Republican content (dark left half of panel Fig.~\ref{fig:exposure_landscape_triple}c). 
Conservatives, on the other hand, report seeing mostly diagonal/ideologically consonant Republican-aligned content, though somewhat less of it than liberals see of Democrats. Instead, their diets incorporate a considerable amount of Democrat-aligned, mid-left/far-left content (relatively bright bottom-right quadrant of panel a)---perhaps reflecting a perception of unavoidable ``liberal mainstream media".

These data are certainly imperfect, not least because they were collected at the end of an already-long ($\approx 25$ minute) survey. However, they allow us to create an estimate of the overall systematic bias of the modern political information environment, a very powerful amalgamation of psychological and algorithmic biases---as dependent on observer's own subjective ideological identity, in the same quantitative domain as the rest of our data. 

\begin{figure*}[htp] 
    \centering
    \begin{overpic}[width=.49\textwidth]{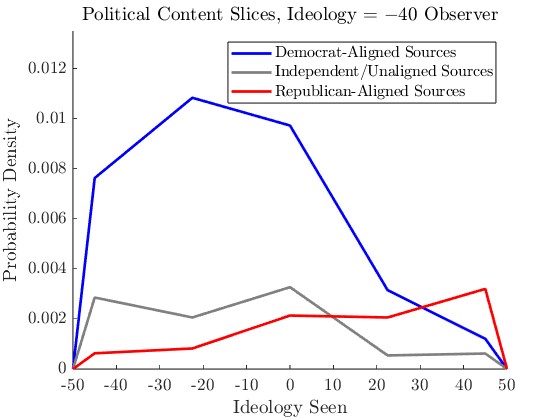}
        \put(2,67){\large{\textbf{a}}}
    \end{overpic}
    \begin{overpic}[width=.49\textwidth]{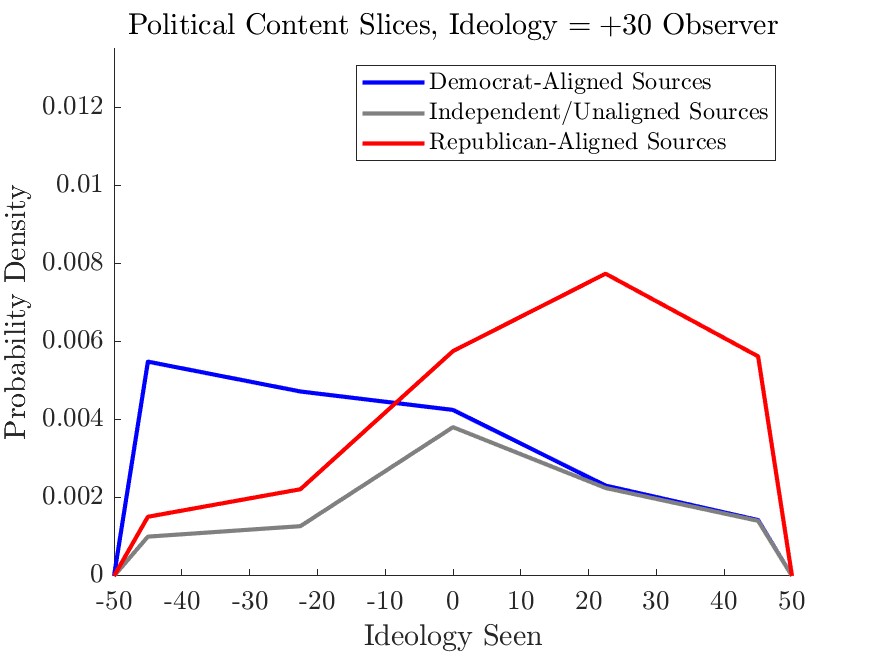}
        \put(1,68){\large{\textbf{b}}}
    \end{overpic}
    \caption{\textbf{Example content diets:} Content exposure curves for individuals at \textbf{a)} $I=-40$ (extremely/very liberal) and \textbf{b)} $I=+30$ (very conservative), taking slices corresponding to their ideology value from the three exposure landscapes in Fig.~\ref{fig:exposure_landscape_triple}. Together, the three types of content add to an area of 1. Visible here is the primary asymmetry: the conservative sees more liberal content than vice-versa. Also, we note that each side sees a more extreme side of the out-group---in \textbf{a}, the red curve peaks at ``far right" (interpreted as +45 here) and in \textbf{b}, the blue curve peaks at ``far left" (-45). These judgments are subjective, so this may be an interpretational difference (e.g.~the same speech could be interpreted as far-right by liberals and mid-right or centrist by conservatives), but that difference is meaningful/operative as far as the impact on the viewer, so these distributions do represent the primary image each side sees of the other---i.e., that the extreme members of the other side are more visible or loom larger in respondents' memory.}
    \label{fig:exposure_distribs_particular}
\end{figure*}

\subsection{Dynamical Modeling}
\subsubsection{Structure}
As mentioned in the Introduction, these two surfaces can drive a dynamical model for popular ideology, with individuals observing a particular slice of political content from the content exposure landscape, agreeing or disagreeing with what they see according to the appropriate probability-distribution slice of the agreement surface, and marginally updating their ideology as a result. 

These models can extrapolate the implications of survey data under various dynamical hypotheses; due to the difficulty of directly measuring ideological drift from individual political stimuli, we are left to theorize how observable quantities (e.g.~agreement and dissonance) come together to contribute to such motion.

For all models to follow, we consider a finite $[-1,1]$ domain for individual ideology $g$ 
and percept ideology $p$ (reflecting the $[-50,50]$ response domain from the survey). To soft-bound dynamics within this domain, we include a damping factor of $(1-g^2)$ on most dynamical effects. This additionally reflects the asymptotic nature of political extremism, where the endpoints are only asymptotically approachable and motion becomes slower as those individuals are more ``entrenched." 

To turn a hypothesized dynamic into a net-drift curve, we integrate over the appropriate joint distribution of $\Delta g$ (a probability of each percept value $p$, representing dissonance $d = p-g$, which engenders a chance of each level of agreement $a$).

\subsubsection{Notation}
We are faced with unfortunate overloading of the natural letters of choice for several quantities: i (ideology, individual/index, or various conventions from physics and math) and p (percept, political party, probability). We will attempt to keep things clear with this section.

We will use subscript $i$ to denote qualities of a particular individual, and subscript $j$ to denote qualities of a particular percept---we avoid ambiguity since this model framework does not include explicit interactions between individuals, rather individuals reacting to percepts drawn from a (systematically biased) environment.  Function inputs will be denoted with square brackets, [ ], to avoid confusion with parentheses. We will use an overscript tilde (e.g., $\Tilde{p}$) to denote the probability density function from which a particular value was drawn (i.e.~$\Tilde{p} = \textrm{Prob}[p]$). For quantities, we have
\begin{itemize}
    \item individual ideology: $g$ (general) or $g_i$ (specific) 
    \item percept ideology: $p$ (general) or $p_j$ (specific)
    \begin{itemize}
        \item percepts are drawn from a distribution $\Tilde{p} = \Tilde{p}[p | g]$, a vertical slice of the surface in Fig.~\ref{fig:exposure_landscape_combined}, or (if percept party matters) $\Tilde{p}_*=\Tilde{p}_*[p|g]$, a vertical slice of a particular party $*$'s surface from Fig.~\ref{fig:exposure_landscape_triple}
    \end{itemize}
    \item dissonance: $d = p - g$  or $d_{ij} = p_j - g_i$ (dissonance of percept $j$ seen by person $i$)
    \item political party: $P_i$ for observer $i$'s party, $P_j$ for percept $j$'s party
    \item agreement $a$ (general) or $a_{ij}$ (the agreement of person $i$ with percept $j$), drawn randomly from distribution $\Tilde{a}[a|d]$, the vertical slice of Fig.~\ref{fig:reaction_surface} for dissonance $d$
    \item time constant $\tau$ (implied, for time-scale setting)
\end{itemize}

\subsubsection{Noise}
Depending on our simulation paradigm, we can have different implementations of ``noise" in the system. 

With a discrete-event-modeling approach, the inherent randomness of events and reactions might provide variance of outcomes. These simulations have the most natural implementation from data, but unfortunately their results depend on the chosen time-step: in particular, if the time-step shrinks, behavior converges towards the mean effect of all content distributions, and all possible reactions to those distributions, acting together proportional to their likelihood---a deterministic result with no variability between individuals, only convergence to fixed points.

If we would rather the results not depend on the simulated time-step, we can take the continuous-time limit of the system---with ``drift" behavior equal to the mean result described in the previous paragraph (the integral of content seen against its possible impacts), and stochastic ``diffusive" behavior determined by constructing a one-dimensional probability distribution of $g'$ at each $g$ value. In particular, we can take slices of the probability space of percept ideology $p$ and agreement $a$ corresponding to level sets of the $g'$ function. If we take the standard deviation of this distribution to be the natural magnitude of the stochastic term (a nontrivial interpretive step we propose in \cite{DSM2024langevin}), we can use standard numerical integration and Monte Carlo simulation of many individuals to examine equilibrium distributions.

We might also impose external noise, perhaps reflecting real-world events (small or large) not captured by our respondents' spontaneous self-recall. We suspect the survey would be an underestimate of influences, considering respondent fatigue and the large number of experiences which contribute to individuals' worldview. However, this was not necessary for any of the simulations to follow.

\subsection*{Model 1: Movement = Agreement $\times$ Dissonance}
\begin{figure*}[htp] 
    \centering
    \includegraphics[width = 0.45\textwidth]{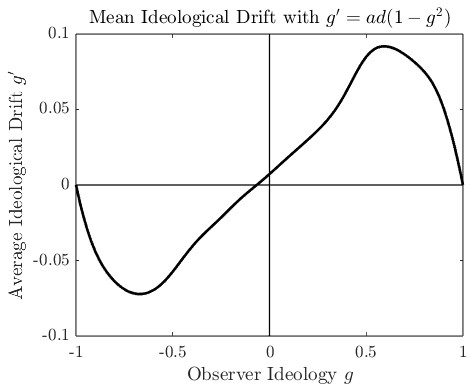} 
    \includegraphics[width = 0.45\textwidth]{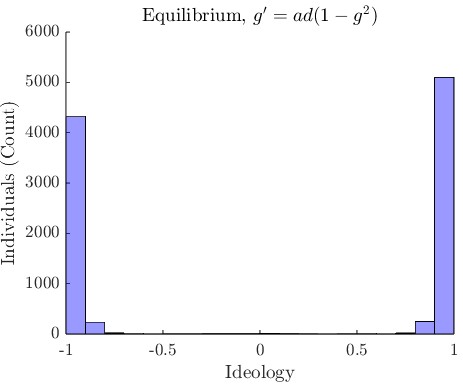} 
    \caption{ Average ideological drift as a function of current ideology according to the most basic dynamical model (\ref{eq:basic_model}). Equilibrium distribution(s) from stochastic numerical integration; noise provided by inherent standard deviation of the distribution of $g'$.
    }
    \label{fig:dynamics_basic}
\end{figure*}

The simplest model of ideological dynamics involving the observable variables is that a percept of dissonance $d$ which engenders agreement $a$ leads to movement $a\cdot d$. Combined with a soft-bounding term, that leads to the model 
\begin{align} 
    g'&= ad (1-g^2) \label{eq:basic_model}
\end{align}
After integrating over the appropriate joint distribution from data, this model implies the net dynamics in Fig.~\ref{fig:dynamics_basic}, suggesting a rather bleak picture of hyperpolarization.

\subsubsection*{Interpretation}
\begin{figure*}[htp] 
    \centering
    \includegraphics[width = 0.48\textwidth]{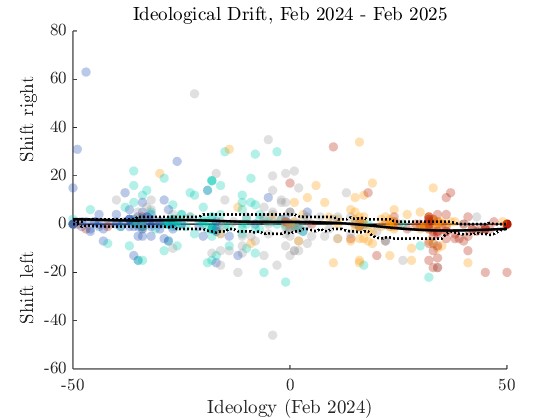} 
    \includegraphics[width = 0.48\textwidth]{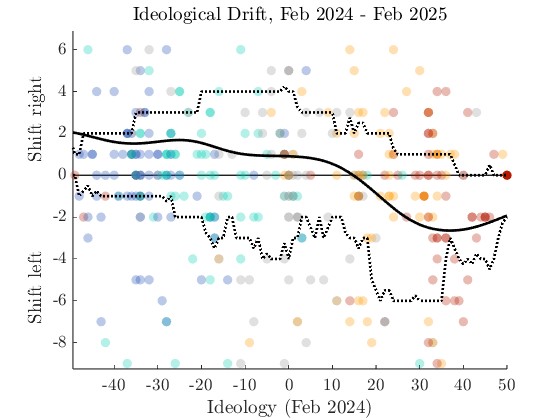} 
    \caption{ \textbf{Empirical self-reported ideological drift (full and zoom):} Participants provided their current (February 2025) self-reported liberal/conservative ideological alignment as well as their recollection of their ideological alignment one year prior (February 2024). The difference between these two serves as a coarse, one-sided approximation of their ideological ``derivative", plotted against their ideology in February 2024 to best match the time of the other data collection for the ``forcing" environment supposedly responsible. \textbf{Solid Curve: } Gaussian-weighted moving mean ($\sigma=8$). \textbf{Dotted curves: } Moving 25/75$^{th}$ percentiles (window width = $15$). Due to the near-zero mean drift, the right panel provides a zoomed view, which shows minor centralizing drift overall---the opposite pattern our basic model suggested.
    }
    \label{fig:empirical_drift}
\end{figure*}

We are faced with two interpretations of the result shown in Fig.~\ref{fig:dynamics_basic}: 
\begin{enumerate}
    \item Our mechanistic model may be correct but dynamics must be very slow (i.e., the U.S.~population as of late 2023/early 2024 was actively polarizing and not at equilibrium with the media environment of the day). 
    \begin{itemize}
        \item In this case, we might test the model's plausibility by comparing this prediction to empirical measures of ideological \textit{velocity} as a function of current ideology. We conducted a brief follow-up survey in February 2025 to test this, with results shown in Fig.~\ref{fig:empirical_drift}---the results did \textit{not} support this polarizing perspective. Recent events at the time of the survey (during the early chaos of the second Trump presidency) could conceivably bias these results towards centralization, however we also asked for participants' estimation of their ideology five years prior (February 2020), and the same pattern (mostly null drift, with slight centralization on average) appears when comparing any two time-slices
        , suggesting that these participants were most polarized in 2020 and have been slowly depolarizing ever since.
    \end{itemize}
    \item We want this model to represent ideological drift on the scale of days, rather than years, and hope for its equilibrium to reflect the current U.S.~ideological distributions. 
    \begin{itemize}
        \item In this case we must augment our theorized dynamics, because Fig.~\ref{fig:dynamics_basic}'s prediction of hyperpolarization to $\pm 1$ does not match the current reality of slightly-overlapping party distributions (as seen in, e.g., \cite{DSM2025spectrograph, pew2017divide, bail2018exposure}).
    \end{itemize}
\end{enumerate} 
We will now explore several augmentations which aim to salve the second interpretation; we seek to uncover plausible day-to-day dynamics which exhibit current ideological distributions as equilibria (after considering the addition of noise).

\subsection*{Model 2: Adding Centralizing Force}
\begin{figure*}[htp] 
    \centering
    \includegraphics[width = 0.45\textwidth]{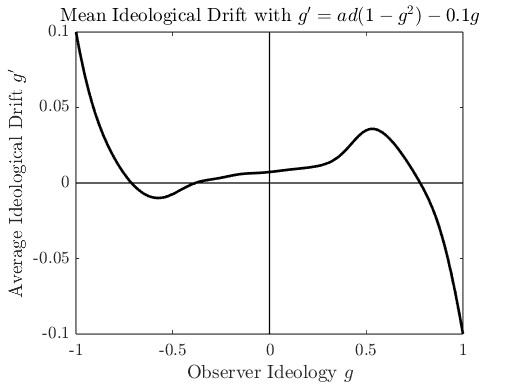} 
    \includegraphics[width = 0.45\textwidth]{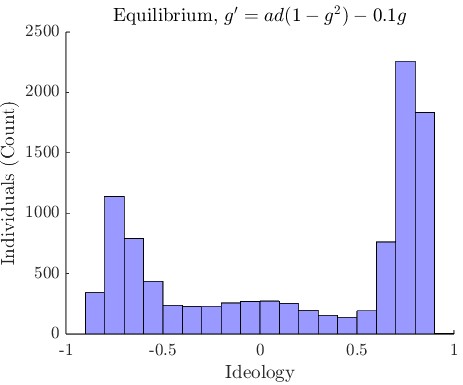} 
    \caption{ Average ideological drift as a function of current ideology with a linear centralizing term (inherent noise only). A bifurcation soon after this value of $k=0.1$ destroys the liberal fixed point and leaves only the conservative camp. 
    }
    \label{fig:dynamics_centralizing}
\end{figure*}

We might hypothesize that there are social costs to being more ideologically extreme, which could manifest as a linear restoring term:
\begin{equation}
    g' = ad(1-g^2) -kg \label{eq:linear_restoring}
\end{equation} 
Note: we leave this outside the otherwise-global damping factor, since we want it to significantly affect people at the edges. This will move the equilibria inwards, for example see Fig.~\ref{fig:dynamics_centralizing}, but is insufficient to create any kind of realistic distributions. 

\subsection*{Model 3:  Saturating Dissonance, Tribalism, Exposure Error, and Party Cohesion}
The strength of high-dissonance disagreement in the basic multiplicative model ends up dwarfing other effects---not only is the magnitude of dissonance directly weighted in the drift magnitude (meaning low-dissonance percepts are naturally diminished in their effect), but faraway percepts are also nearly certain to cause strong disagreement (while low-dissonant percepts cause only mild agreement). There are several mollifying fixes we will implement here and hopefully justify, but others are certainly possible.\\

\uline{\textit{Saturating Dissonance.}}
First, we suppose that beyond a certain point, individuals are less sensitive to changes in dissonance. This might be plausibly represented with any sigmoid function, but as an initial example we will have the effect of dissonance simply saturate at $\pm1$. We use the notation 
\begin{align*}
    \hat{d} = \textrm{min}(\textrm{max}(d,-1),1)
\end{align*}
for ``capped-magnitude" dissonance.\\

\uline{\textit{In-Group Favoritism.}}
While our survey was unable to sufficiently quantify a partisan/tribalist bias in agreement (due to in-group/out-group interactions inhabiting different parts of the dissonance domain, as seen in Fig.~\ref{fig:reaction_curve_ingroup_outgroup}) or content exposure (due to observers of different parties exhibiting so little overlap in the ideology domain), any model which fails to incorporate political party in some manner will not be able to exhibit the slightly-overlapping unimodal distributions per party that we see in reality (in, e.g.,~\cite{pew2017divide, bail2018exposure, DSM2025spectrograph}). For a basic implementation of partisan tribal bias, we will up-shift effective agreement for in-group percepts by 0.3 and those that are only one party-step away by 0.15:
\begin{align*}
    \hat{a}_{ij} &= a + 0.15*(2-|i-j|)
\end{align*}
where $i,j \in {-1,0,1}$ corresponding to Democrat, Independent, and Republican observers/content respectively. That is, when an individual perceives an in-group percept, their break-even point for ideological movement away from the source is $-30$ agreement rather than $0$, i.e., one must disagree quite strongly with their in-group to be repelled.\\

\textit{\uline{Exposure Estimation Error/Out-Group Scaling.}} 
Taking the data at face value, liberals are barely influenced by Republican-aligned percepts and conservatives are predominantly influenced by Democrat-aligned ones. However, we must consider that the estimation of exposure to each ``allegiance" of political content was a coarse and possibly confusing part of our survey, and might itself be subject to significant and systematic inaccuracy. 

If we consider this asymmetry partially a survey-interpretational difference, perhaps that liberals underestimate how much they see of Republican politicians (which might engage with out-group bias and could therefore be better classified as Republican-aligned content, but interpreted as Democrat-aligned if it was discussed on a Democrat-aligned program), and that conservatives over-estimate the amount they see from Democrat-aligned sources (perhaps reflecting the reverse effect, overestimating how much their content is dominated by the Democratic-affiliated sources). Whatever the reasoning, until better perception data is gathered---with a clearer definition of what counts as an out-group percept---it is tempting to allow some wiggle room on the two cross-party exposure rates, or equivalently, cross-party influence magnitudes. To create a realistic-looking equilibrium from this initial data, we now introduce two more parameters---our first and only asymmetric ones---and suppose that Democrats are 40\% more affected by out-group content than they claim, and that Republicans are 45\% \textit{less} affected by out-group content than they claim. That is, we substitute 
\begin{align*}
        g'_{DR} &\rightarrow 1.4\ g'_{DR} \nonumber \\
        g'_{RD} &\rightarrow 0.55\ g'_{RD}
\end{align*}

\textit{\uline{Party Cohesion.} }
Finally, we might suppose that there are some conformity-fostering incentives which manifest as a cohesive force pulling individuals towards their current party mean. This is the first true individual-to-individual coupling in the model, and as such complicates any theoretical analysis of the model immensely as the current state of the population alters the forcing landscape; but for practical purposes (and at small magnitude) it merely smooths out any otherwise-multimodal party distributions (namely the Democrats, which otherwise have distinct radical and central contingents). 

\begin{align}
    g'_{coh} = 0.03(\mu_i - g)(1-g^2),
\end{align}
where $\mu_i$ is the mean ideology of individual $i$'s party.\\

\begin{figure*}[htp] 
    \centering
    \includegraphics[width = 0.4615\textwidth]{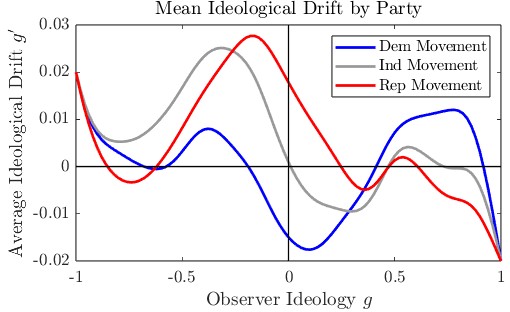} 
    \includegraphics[width = 0.451\textwidth]{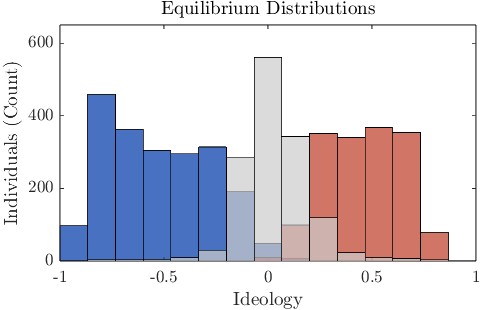} 
    \caption{ Average ideological drift (excluding cohesion, which depends on the mean of each simulated population) as a function of current ideology for Model 3, with a centralizing term, capped effective dissonance, basic positive tribalism, and exposure-asymmetry reduction. The equilibrium distributions in the right panel represent 10,000 simulated individuals: (4,000 Democrat, 2,000 Independent, and 4,000 Republican) who were initialized at ideologically consonant normal distributions and run until equilibration. In the left panel, the right side of the blue curve and left side of the red curve are not well informed by data (we had no respondents of those types in those ideology regions), and thankfully remain unused unless individuals start in those anomalous regions.
    }
    \label{fig:dynamics_multiEffects}
\end{figure*}

\noindent \textbf{\uline{Full Candidate Model.} }
All together (and still including a slight centralizing bias 
in the same manner as Model 2), we have 
\begin{align}
    g_{ij}' = (\hat{a}_{ij}&\hat{d} + 0.03(\mu_i - g))(1-g^2) - 0.02 g \label{eq:multi_effects} \\ 
    & \qquad \textrm{with } \begin{cases}
        g'_{DR} \rightarrow 1.4\ g'_{DR} \nonumber \\
        g'_{RD} \rightarrow 0.55\ g'_{RD}
    \end{cases}\\
    \delta(g,i) &= \sum_j P(g'_{ij}(g)) \nonumber \\
    F(g,i) &= \textrm{mean}(\delta(g,i)) \nonumber \\
    G(g,i) &= \textrm{std}(\delta(g,i)) \nonumber \\
    \textrm{d}g(g,i) &= F(g,i) \textrm{d}t + G(g,i) \textrm{d}W\nonumber 
\end{align}
which results the dynamics and equilibria in Fig.~\ref{fig:dynamics_multiEffects}. 

So with six parameters and a few simple functional assumptions we have a \textit{plausible} model of ideological dynamics which turns the observed micro-scale measurable quantities into (a rough version of) the observed macro-scale distributions. However, the obvious modeling freedom and introduction of so many free parameters is undesirable, and ideally these effects would be quantitatively explored in dedicated experiments. This process of progressively cornering uncertainty and conjecture by constraining with experimental data is a virtuous cycle which has led to progress in many scientific fields, so we are optimistic about its application to this endeavor going forwards.

\section{Limitations}
The modeling framework (which these models all belong to) attempts to strike a fruitful balance between capturing nuance and being informable with realistically attainable and relatively unproblematic data. There are several important implicit concessions this framework makes, which may be fixable with improved data and cleverer modeling in the future.

First, our assumption of universality/interchangeability of individuals results in the implication that individuals drift ergodically around their party's ideological distribution (i.e., the distribution of individuals at equilibrium serves as a probability distribution for any single individual's position over time), a quality unlikely to be borne out by empirical longitudinal data of real people. As mentioned in the ``Other Possible Models" section, it may be possible to approximate the ``stickiness" that personal convictions and psychological proclivities contribute to individuals' ideology with a decaying ``temperature" variable for each individual which damps movement over time, but assigning actual inherent psychological differences is a proposition notoriously fraught with possible experimenter bias and other social-science challenges.

Second, our continuous-time drift (while convenient to avoid consideration of individualized political interaction frequency) does not allow large discontinuous jumps which might occur alongside major recontextualization events such as, e.g., being the victim of a crime or a family member coming out as gay. A discrete-time analogue may be easily implemented (using individual random events with randomly-drawn agreement values), but large-deviation events may be hard to govern quantitatively due to the unpredictability of these real-world scenarios.

Third, the finite one-dimensional domain does not reflect possible psychological similarities between more extreme ideologues on both ends of the spectrum \cite{deBruin2025extremesSimilar} 
which may lead individuals to jump from one side to the other without passing continuously through moderate centrism on the way.  We might view this as just another form of ``individual differences"---that some individuals may be attracted to more essentialist/dogmatic logic and thus \textit{both} outer portions of the spectrum (as theoretically, these mindsets are closer to each other than they are to centrism in some higher-dimensional representation of political reasoning space which is the true attractor). 

These complex psychological realities present fertile ground for future work. While most frontiers will doubtless resist explicit quantitative ``solving" as they have for centuries, we hope that nuanced quantitative approaches can provide a scaffolding around which scientific data-based consensus can gradually be discovered.



\section{Conclusions}
Recent data shows that political reasoning among U.S.~respondents has a consistent core trend, whereby acceptance of political positions falls systematically with distance on the subjectively-perceived one-dimensional ``political spectrum.''  The robustness and symmetry of this response seems to point to a universal underlying reasoning mode, at least in the context of a two-party ``us and them''-type system.


The linking of these findings within the original mathematical framework allows for the systematic exploration of how individual reaction tendencies and environmental input biases combine, creating population-level dynamical outcomes such as polarization. 


We have presented the implications of recent data under a dynamical framework. With the simplest interpretation of the effect of ``agreeing'' with content, we arrive at an ideological forcing function which indicates increasing polarization, with the unstable indifference point slightly right of center. 

We might interpret this as the true forcing on the population, but on a slower time-scale than our simulations. 

Alternately, we may search for more complex ideological dynamics which might turn observed data on perceptions and agreement into observed distributions. These additional theoretical components included capping dissonance (corresponding to a maximum ``otherness'' beyond which the observer's psychological response is identical), adding a blanket centralizing force (representing the social penalties of being politically extreme in everyday life, and general benefits of normalcy/conformity), 
positively shifting agreement (reflecting that either individuals have a negative response bias when reporting their own feelings about politics, or that the neutral-movement point occurs with slight disagreement).


\section{Methods}


Many details of the survey methodology have been discussed in detail in previous work \cite{DSM2025spectrograph}. However, this work differs slightly in its inclusion criteria. 

The survey was administered to Mechanical Turk Masters (N = 166), volunteers (N = 130), and a representative sample of respondents from the Prolific platform (N = 508). For the purpose of creating the most robust and reliable models possible, the curves and surfaces in this paper use all survey samples combined---each respondent only serves as a ``representative" of their own self-professed ideology, and everything is normalized so that representativeness of the U.S.~population is not a concern. The observed patterns are consistent across sub-populations (data available upon request), but converge to the smoothest overall result with all data included.

\subsection{Data cleaning protocol}
 
Also in the interest of improving the resulting model, an effort was made to filter out unreliable/un-serious responses with a data-cleaning protocol, as there were a small but noticeable number of nonsensical outliers (from botting/rushing/non-comprehension/trolling). For each suspicious signature (criteria below) respondents generate a suspicion score, and those accumulating 5 or more of these red flags were filtered out.  

The red-flag criteria were:
\begin{itemize}
    \item very large ($>$20) shifts in ideology between the start and end of the experiment (1 point)
    \item severe misalignment of party identification with ideology (Strong Democrats rightward of +20, Lean Democrats rightward of +40, and vice-versa for Republicans) (2 points)
    \item rating individual liberal-pool statements as \textit{extremely} conservative ($>$40) or vice-versa (1 point each)
    \item rating an entire liberal/conservative pool on average as contrary to their usual side of the spectrum (1 point per 10 ideology points wrong side of 0)
    \item ``strong'' partisans disagreeing on average with their chosen side's statement pool (1 point)
    \item having a standard deviation of all responses (on all sliders) below 15---indicating spam-clicking without consideration---(1 point for each 4 the stdev is less than 15)
    \item claiming to be near-perfectly evenly exposed to all 15 types of content (including, e.g., far-right Democrat-aligned and far-left Republican-aligned) (1 point)
    \item having a large fraction of extreme endpoint positioning of all sliders (1 point per 15\% of responses)
\end{itemize}
Overall, 28 out of 804 (3.5\%) respondents were filtered out under these conditions, leaving 776 (96.5\%) to inform our surfaces. Raw-data versions of these surfaces (as well as the responses themselves) are available upon request.

\begin{acknowledgments}
The authors thank Mary McGrath for assistance with the survey design and financial backing for the Prolific survey, and Marisa Eisenberg for advice and guidance. We also thank Drew Trygstad for his help with formatting the survey, and Daniel Abrams for his assistance with the original mathematical model and early conceptual stages of the survey. Additionally, the authors' sincere gratitude goes out to all volunteers who offered their time and attention to contribute to this endeavor.
\end{acknowledgments}

\bibliography{refs}

\end{document}